\documentclass[utf8]{FrontiersinHarvard} 

\usepackage{url}

\definecolor{xlinkcolor}{cmyk}{1,1,0,0}
\usepackage[
  bookmarks=true,         
  pdfnewwindow=true,      
  colorlinks=true,        
  linkcolor=xlinkcolor,   
  citecolor=xlinkcolor,   
  filecolor=xlinkcolor,   
  urlcolor=xlinkcolor,    
  final=true,
]{hyperref}

\usepackage{microtype,subcaption}
\usepackage[onehalfspacing]{setspace}

\usepackage{newtxtext}
\usepackage[smallerops,cmbraces,varvw]{newtxmath}
\usepackage[normalem]{ulem}
\usepackage{csquotes}

\newcommand*{\OIII}{[O\,\textsc{iii}]} 
\newcommand*{\Hbeta}{\ensuremath{\mathrm{H}\beta}} 
\newcommand*{\Msun}{\ensuremath{\mathrm{M}_\odot}} 
\newcommand*{\ngc}[1]{NGC\,#1}
\newcommand*{\m}[1]{M\,#1}

\usepackage{siunitx} 
\DeclareSIUnit[number-unit-product = ]\percent{\char`\%} 
\DeclareSIUnit\mag{mag}
\DeclareSIUnit\year{yr}
\DeclareSIUnit\Msun{M_\odot}

\usepackage[capitalize]{cleveref} 
\crefname{section}{Sect.}{Sects.}
\creflabelformat{equation}{#2#1#3} 

\usepackage{aas_macros}

\def\keyFont{\fontsize{8}{11}\helveticabold }
\def\firstAuthorLast{Valenzuela {et~al.}} 
\def\Authors{Lucas M. Valenzuela\,$^{1,*}$, George H. Jacoby\,$^{2}$, Rhea-Silvia Remus\,$^{1}$, Marcelo M. Miller Bertolami\,$^{3,4}$ and Roberto H. Méndez\,$^{5}$}
\def\Address{$^{1}$Universitäts-Sternwarte, Fakultät für Physik, Ludwig-Maximilians-Universität München, Scheinerstr. 1, 81679 München, Germany\label{inst:usm}\\
$^{2}$NSF’s NOIRLab, 950 N. Cherry Ave., Tucson, AZ 85719, USA\label{inst:noirlab}\\
$^{3}$Instituto de Astrofísica de La Plata, UNLP-CONICET, La Plata, Paseo del Bosque s/n, B1900FWA, Argentina\label{inst:ialp}\\
$^{4}$Facultad de Ciencias Astronómicas y Geofísicas, UNLP, La Plata, Paseo del Bosque s/n, B1900FWA, Argentina\label{inst:unlp}\\
$^{5}$Institute for Astronomy, University of Hawaii, 2680 Woodlawn Drive, Honolulu, HI 96822, USA\label{inst:ifa}}
\def\corrAuthor{Lucas M. Valenzuela}

\def\corrEmail{lval@usm.lmu.de}

\begin{document}
\onecolumn
\firstpage{1}

\title[PICS II: Circumnebular extinction effects on the PNLF]{The PICS Project: II. Circumnebular extinction variations and their effect on the planetary nebula luminosity function} 

\author[\firstAuthorLast ]{\Authors} 
\address{} 
\correspondance{} 

\extraAuth{}

\maketitle

\begin{abstract}

\section{}
For decades, the theoretical understanding of planetary nebulae (PNe) has remained in tension with the observed universal bright-end cutoff of the PN luminosity function (PNLF).
The brightest younger PN populations have been observed to be fainter in their \OIII{} emission than expected. Recent studies have proposed that circumnebular extinction is a key ingredient in bringing their brightness down to the observed level.
In this work we use the recently introduced PICS (PNe In Cosmological Simulations) framework to investigate the impact of different circumnebular extinction treatments on the modeled PNe and their PNLF for a large range of stellar ages and metallicities.
We test how different slopes in the observed relation of extinction versus central star mass modify the bright-end cutoffs of the PNLF, finding that steeper slopes lead to large changes for young stellar populations. In contrast, the differences for older PNe are much smaller.
However, for individual PNe, the extinctions observed in nearby galaxies appear to be much higher than the models predict, showing that improvements on both the modeling and observational sides are needed to gain a better understanding of the brightest and strongly extincted PNe.
These findings further advance the theoretical foundation for interpreting observed extragalactic PN populations coming from more complex composite stellar populations in the future.

\tiny
 \keyFont{ \section{Keywords:} planetary nebulae, luminosity function, circumnebular dust extinction, modeling, parameter testing} 
\end{abstract}

\section{Introduction}
\label{sec:introduction}

The bright-end cutoff of the planetary nebula luminosity function (PNLF) has long been used as a standard candle in the cosmic distance ladder \citep[e.g.,][]{jacoby89:pnlfI, rekola+05}.
This has been based on the empirical finding that the brightest planetary nebulae (PNe) in a given galaxy have the same \OIII{} $\lambda5007$ emission line absolute magnitude of $M^*(5007) = -4.5$, independent of galaxy morphology \citep[e.g.,][]{ciardullo+02:pnlfXII}. This universal PNLF bright-end cutoff appears to be independent of stellar age and metallicity above $12 + [\mathrm{O}/\mathrm{H}] \gtrsim 8.5$ \citep[e.g.,][]{ciardullo12}.
This universality has stood in contrast with theoretical expectations for a long time, where younger stellar populations with more massive PN central stars should feature much brighter PNe than older, less massive central stars \citep[e.g.,][]{ciardullo22}.

One proposal for bringing the brightness down for PNe in younger stellar populations has been the self-extinction through the dust formed in the winds during the asymptotic giant branch (AGB) phase preceding the PN phase \citep[e.g.,][]{ciardullo&jacoby99, ventura+14}.
Based on their findings and a model, \citet{jacoby&ciardullo25} showed how this circumnebular extinction seems to exactly compensate the increasing potential brightness of PNe with more massive central stars over a large mass range, which could in part explain the constant PNLF bright-end cutoff.
Considering the brightest PNe in the bulge of \m{31}, \citet{davis+18} found very large circumnebular extinctions, implying intrinsic brightnesses well beyond the expectation for an older stellar population. Their results are also consistent with studies of other nearby galaxies \citep[e.g.,][]{mendez+05, reid&parker10}.
Furthermore, circumnebular extinction has been suggested as a way to understand the connection between UV measurements of post-AGB stars and PN central star candidates \citep{sarzi+11}.
The findings demonstrate that understanding the effects of circumnebular extinction on the bright end of the PNLF and the individual PNe is crucial to properly interpret the observations in both local and distant galaxies.

Recently, theoretical modeling of PNe has made large advances through the novel PICS (PNe In Cosmological Simulations) framework \citep{valenzuela+25:picsI, valenzuela+25:pics}, a model that traces all the properties necessary to obtain a realistic PN population for any given single stellar population (SSP).
Built in a modular fashion, it allows one to study precisely how variations of the models and parameters impact the resulting PN and PNLF properties for any stellar population.

In this work, we use PICS to investigate how the chosen circumnebular extinction recipe affects the bright-end cutoff of the PNLF of different SSPs and compare the individual modeled extinctions of the bright PNe to observations. In \cref{sec:method} we give a brief overview of the PICS model and the extinction recipes tested. We then present the results in \cref{sec:results} and discuss them in \cref{sec:discussion}.

\section{Method}
\label{sec:method}

We use the fiducial model of the PICS framework presented by \citet{valenzuela+25:picsI}, for which we test how the resulting PN properties change based on variations of the circumnebular extinction model.
In the following, we give a brief overview of PICS and the circumnebular extinction recipe variations we tested.

\subsection{PICS model}
\label{sec:pics_model}

The PICS framework is composed of several modules that for a given SSP predict the PN population found therein.
Such an SSP is parameterized by its total mass, age, metallicity, and initial mass function (IMF).
In the PICS fiducial model, these properties are used to first determine the initial mass of the central stars of the current PNe through the lifetime function \citep{miller_bertolami16}.
The final stellar mass is obtained from the initial-to-final mass relation (IFMR) and the stellar properties from the post-AGB stellar tracks from \citet{miller_bertolami16}. All three of these steps are metallicity-dependent.
Based on the stellar properties, the nebular model by \citet{valenzuela+19} is applied together with a metallicity correction from \citet{dopita+92}. The nebular model also accounts for the possibility of the nebula being optically thin to the ionizing photons.
The primary resulting quantity is the \OIII{} flux, $F(5007)$.

As shown by \citet{jacoby&ciardullo25}, the \OIII{} flux can be significantly extincted by circumnebular dust that was expelled through the AGB winds. The extinction is expected to decrease over time during the post-AGB phase as the dust and nebula diffuse into the interstellar medium (ISM) with age.
The fiducial PICS model accounts for a post-AGB time-dependent circumnebular extinction consistent with the empirical findings from \citet{jacoby&ciardullo25}, where we define the post-AGB time as the time after the central star reaches an effective temperature of $T_\mathrm{eff} = \SI{25000}{\kelvin}$.
The implemented extinction recipe is solely a function of post-AGB time of the central star, which in particular means we here assume it is independent of stellar mass.

To derive this equation, an intermediate step has to be taken because the relations from \citet[table~2]{jacoby&ciardullo25} between stellar final mass and extinction, $c(\Hbeta)$, are derived only from the brightest PNe.
For this, we determined the post-AGB time at which a central star (with a metallicity of $Z=0.01$) of given final mass reaches its brightest intrinsic \OIII{} magnitude.
This will be needed to derive an \enquote{equivalent brightest PN final mass}.
The relation by \citet{jacoby&ciardullo25} returns an extinction based on the final mass of a brightest PN. From these two relations, we derived a direct dependence of extinction on post-AGB time: From the post-AGB time we obtain the \enquote{equivalent brightest PN final mass}, from which we finally get the extinction.
Because we make the simplifying assumption that the extinction depends only on post-AGB time and not on stellar mass, we use the resulting relation for all PNe.
For the fiducial PICS model, we use their orthonormal regression fit to the combined Large Magellanic Cloud (LMC) and \m{31} sample of PNe, the \emph{Combined OR} (orthogonal regression) fit (from the table of \citealp{jacoby&ciardullo25}).

By running PICS for a given SSP, the result is a population of PNe, each with an \OIII{} magnitude.
The resulting PNe differ from each other because the nebular model by \citet{valenzuela+19} is stochastic in nature. Two reasons for this are the randomly drawn post-AGB lifetime and the randomness of the optical thickness of the nebula.
In this work, we leave the fiducial PICS model fixed, with exception of the circumnebular extinction treatment. The variations thereof are described in the following.

\subsection{Circumnebular extinction variations}

While we used the Combined OR fit of \citet{jacoby&ciardullo25} for the fiducial PICS model, they also provided five additional fits to their PN data, varying both the subsample of PNe (Combined, LMC, \m{31}) and the fitting method (OR; ordinary least squares, OLS). They also include a fit that optimizes the final mass range over which the bright-end cutoff magnitude remains constant at $M^*(5007) \sim -4.5$ (named \emph{Optimized}, see more details in Sect.~5 of \citealp{jacoby&ciardullo25}).

For the Combined OR fit, we assumed a metallicity of $Z=0.01$ for the post-AGB evolution needed for the intermediate step (see the previous section). However, for the pure \m{31} and LMC fits, it is more appropriate to assume metallicities that would more closely match those expected for their respective brightest PNe. Only for the Combined and Optimized fits did we assume $Z=0.01$.
For the LMC fits, we used $Z=0.007$ based on the upper metallicity found for the LMC and its PNe \citep[e.g.,][]{dopita&meatheringham91:II}.
For \m{31}, we used a solar-like metallicity of $Z=0.015$, based on metallicity measurements of luminous PNe in \m{31} \citep[e.g.,][]{balick+13}.
For each of the metallicities, we fit a broken linear fit (above and below $M_\mathrm{final} = \SI{0.60}{\Msun}$) to the relation of final mass versus post-AGB time at which an optically thick PN reaches its brightest \OIII{} luminosity, analogously to the $Z=0.01$ fit by \citet[appendix~D]{valenzuela+25:picsI}.
The derived relations for the two metallicities are the following:
\begin{equation}
    M_\mathrm{final,bright}^{Z=0.007} / \Msun = \max \left\{ \begin{aligned}
        1.023 - 0.1740 \log(t_\mathrm{post-AGB} / \si{\year}) \\
        0.714 - 0.0432 \log (t_\mathrm{post-AGB} / \si{\year})
    \end{aligned} \right\},
    \label{eq:brightest_ages_LMC}
\end{equation}
\begin{equation}
    M_\mathrm{final,bright}^{Z=0.015} / \Msun = \max \left\{ \begin{aligned}
        0.956 - 0.1477 \log(t_\mathrm{post-AGB} / \si{\year}) \\
        0.743 - 0.0531 \log (t_\mathrm{post-AGB} / \si{\year})
    \end{aligned} \right\},
    \label{eq:brightest_ages_M31}
\end{equation}
where the respective maximum of the two linear terms is taken.
Based on the fit parameters presented in table~2 of \citet{jacoby&ciardullo25}, the extinction $c(\Hbeta)$ is finally determined according to
\begin{equation}
    c(\Hbeta) = \mathrm{S} \times M_\mathrm{final,bright} + I,
\end{equation}
where $S$ is the slope and $I$ the intercept.
One important note is that $M_\mathrm{final,bright}$ is not to be confused with the actual final mass of a PN. It is rather the \enquote{equivalent brightest PN final mass} as described in \cref{sec:pics_model}, based on the post-AGB time of a given PN (which can have any other final mass), and needed as input for the fits by \citet{jacoby&ciardullo25}.

In addition to the relations taken directly from \citet{jacoby&ciardullo25}, we also considered a shift of the OR extinction relations of $\Delta c(\Hbeta) = +0.08$.
The motivation for this shift arises in \cref{sec:results_chbeta} and is explained there in detail.
Finally, the extinction in \Hbeta{} is converted to $A_{5007}$ according to eq.~13 of \citet{valenzuela+25:picsI}.

\begin{figure}[!t]
    \begin{center}
    \includegraphics[width=0.6\textwidth]{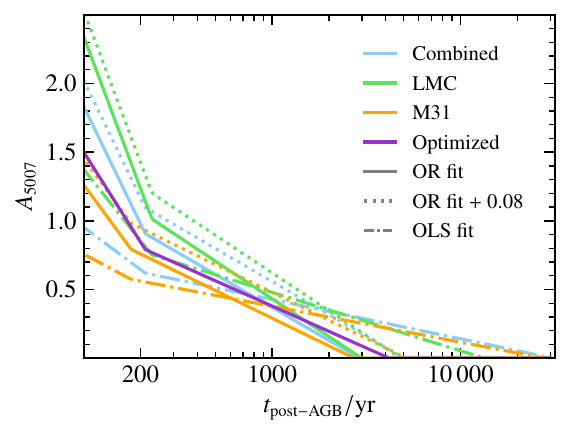}
    \end{center}
    \caption{Circumnebular extinction of \OIII{}, $A_{5007}$, as a function of post-AGB time of the central star for the different extinction recipes tested. The recipes are based on the relations by \citet{jacoby&ciardullo25} for the extinction $c(\Hbeta)$ versus core mass for the brightest PNe in \m{31} and the LMC. The colors indicate the Combined (blue), LMC (green), \m{31} (orange), and Optimized (purple) fits, and the line styles denote the respective OR (solid), shifted OR (dotted), and OLS (dash-dotted) fits for all but Optimized. The dotted \emph{OR fit + 0.08} lines are based on their OR relations shifted by $\Delta c(\Hbeta) = +0.08$.}
    \label{fig:a5007}
\end{figure}

The result is an \OIII{} extinction, $A_{5007}$, as a function of post-AGB time, which we show in \cref{fig:a5007}.
All recipes produce broken linear relations of $A_{5007}$ with respect to the logarithmic post-AGB time and have the break at around \SI{200}{\year}.
The OR models have a significantly higher extinction than their OLS counterparts at young ages, but in turn have a slightly lower extinction at ages beyond ${\sim}\SI{1500}{\year}$.
The modified OR models with the $+0.08$ shift have a slight offset to larger extinction by $\Delta A_{5007} = +0.18$.
Between the different underlying observed PN samples, the LMC relations have the most extinction until about \SI{1000}{\year}, with the LMC OLS relation being similar to the \m{31} OR fit. The Combined relations respectively lie in between the LMC and \m{31} relations, but align more closely with \m{31} at later times. The Optimized relation leads to slightly stronger extinctions than the \m{31} OR model.

These variations of the circumnebular extinction recipe were implemented within the PICS framework as described. They are used in the following to investigate their effect on the resulting PN population properties.

\section{Results}
\label{sec:results}

While the recipe for extinction was derived from the maximal brightness of PNe with a given final stellar mass, the nebular model by \citet{valenzuela+19} allows for escaping ionizing fractions and different \OIII{} to \Hbeta{} flux ratios. This leads to the possibility of dimmer PNe arising, leading to a range of possible \OIII{} magnitudes for a given final mass and post-AGB time.

\subsection{Extinction of \texorpdfstring{$\Hbeta$}{Hβ}}
\label{sec:results_chbeta}

\begin{figure}[!t]
    \begin{center}
    \includegraphics[width=0.6\textwidth]{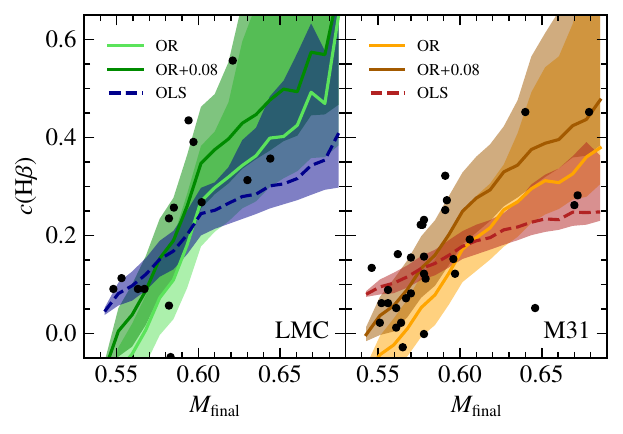}
    \end{center}
    \caption{Circumnebular extinction of \Hbeta{}, $c(\Hbeta)$, as a function of stellar final mass for the brightest observed and modeled PNe assuming different extinction recipes. LMC-like PNe ($Z=0.007$) are shown on the left and \m{31}-like PNe ($Z=0.015$) on the right. Bright PNe are selected according to $M(5007) \leq -3.5$, analogously to the observed PNe (black circles) as selected by \citet{jacoby&ciardullo25}. The lines indicate the medians and the shaded regions the $1\sigma$-scatter of the extinction values at the respective final masses. The extinction recipes correspond to the input fit relations from \citet{jacoby&ciardullo25}, orthonormal regression (OR), shifted OR by $\Delta c(\Hbeta) = +0.08$, and ordinary least squares (OLS).}
    \label{fig:cHbeta_mfinal}
\end{figure}

As the recipe for circumnebular extinction was informed through the observations by \citet{jacoby&ciardullo25}, this allows for a direct consistency check for the actually derived extinction values of the resulting PN population from the PICS model.
To this end we ran PICS for an LMC-like PN population of $Z=0.007$, for which we only kept those with extincted magnitudes $M(5007) \leq -3.5$ for comparison with the observational sample. We ran PICS with the extinction recipes based on the LMC OR, OLS, and the shifted OR relations.
The $c(\Hbeta)$ extinctions of the bright PNe are shown in the left panel of \cref{fig:cHbeta_mfinal}, where the lines indicate the median extinctions found for the bright PNe at the individual final masses, and the shaded regions the $1\sigma$-scatter. The observed extinctions of the LMC PNe are indicated with black points.
The same is shown for an \m{31}-like PN population of $Z=0.015$ in the right panel for the analogous three \m{31} extinction fits to the observed \m{31} PNe.

For both PN populations, it can be seen that the OR model (light green and light orange) leads to overall lower extinctions of the PNe compared to the observations, though the slope stays the same. In contrast, the OLS model (dashed blue and red) results in too much extinction at lower masses (${\lesssim}\SI{0.56}{\Msun}$) and underestimates the extinction at higher masses (${\gtrsim}\SI{0.59}{\Msun}$) because of its overall more shallow slope.

We find that the OR models lead to systematically too low values of $c(\Hbeta)$ by around 0.08. For this reason, we implemented a shifted OR relation with $\Delta c(\Hbeta) = +0.08$. For the shifted models the PN extinction distributions are indeed brought into much better agreement with the observations (dark solid lines in \cref{fig:cHbeta_mfinal}).

The motivation for such a shift is that the observations on which the relations are based are of the brightest PNe within one magnitude of the bright-end cutoff, and not only of the very brightest PNe.
In contrast, our implemented recipe assumes the extinction from the fits of \citet{jacoby&ciardullo25} to correspond to the point in time where the PNe are at their brightest.
After the time of reaching its brightest \OIII{} flux, a PN drops in brightness and the extinction declines. Such PNe could still reside in the brightest magnitude, but would have a lower extinction.
For this reason it is expected that the average extinction of the brightest PNe ($M(5007) \leq -3.5$) will be underestimated through the application in the PICS model, suggesting that a correcting shift is necessary.
We conclude that the shifted OR recipe leads to the most compatible extinctions with observations and that the shift is reasonable for adaptation within the PICS framework.

\subsection{Extinction effects on the PNLF}
\label{sec:pnlfs}

\begin{figure}[!h]
    \begin{center}
    \includegraphics[width=0.82\textwidth]{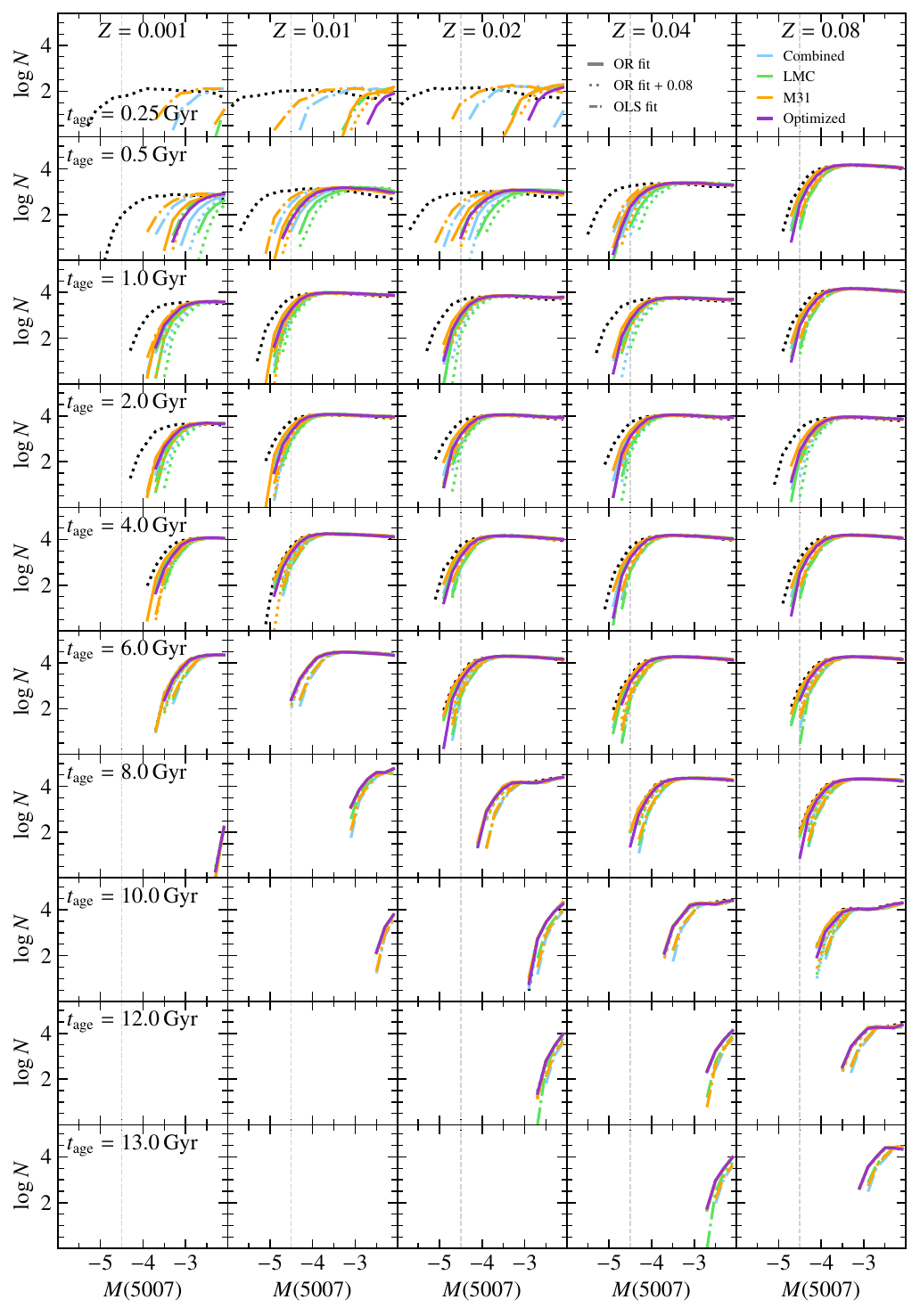}
    \end{center}
    \caption{Grid of SSP PNLFs for different metallicities (columns; $Z=0.001$, 0.01, 0.02, 0.04, 0.08) and ages (rows; $t_\mathrm{age} = 0.25$ to \SI{13}{\giga\year} for the different circumnebular extinction recipes tested (colored lines), as well as without extinction (dashed black lines). The observed universal bright-end cutoff of $M(5007) = -4.5$ is indicated by the dashed gray lines.}
    \label{fig:pnlf_grid_extinction}
\end{figure}

Having verified the agreement of the modeled circumnebular extinction with the original observations, we next tested how much the variations of the extinction recipe actually affect the PNLF and its bright end.
Running PICS for a sufficiently large number of PNe that arise from an SSP of certain age and metallicity produces the PNLF expected for such an SSP.
\Cref{fig:pnlf_grid_extinction} shows a grid of such SSP PNLFs for five different metallicities from $Z=0.001$ to 0.08 and ten SSP ages between 0.25 and \SI{13}{\giga\year}, using no circumnebular extinction (dotted black lines) and the different extinction recipes as described in \cref{sec:method} (solid lines for OR, dotted for shifted OR, and dash-dotted for OLS).

As expected, the extinction is the highest for the young SSPs with high-mass PN central stars, leading to dimmer bright-end cutoffs for the extincted PNe (colored lines) compared to the non-extincted PNe (dotted black lines) in the upper rows.
Here it becomes clear that the OLS recipes lead to the smallest amount of dimming (dash-dotted lines) for young SSPs, followed by the OR (solid lines) and shifted OR recipes (dotted colored lines).
Compared to the Combined OR recipe used in the fiducial PICS model, most extinction recipes lead to brighter PNLFs, except for the two LMC OR recipes. However, these differences become noticeably smaller for ages of \SI{1}{\giga\year} and older, reducing the differences between the circumnebular extinction recipes to ${\lesssim}\SI{0.5}{\mag}$.
These differences between the tested recipes remain rather constant for the metallicity range considered ($Z = 0.001$--0.08) and also moving towards older SSPs. For the oldest SSPs of $t_\mathrm{age} \gtrsim \SI{8}{\giga\year}$, the dimming becomes zero or near-zero for all but the OLS recipes. This is related to the more shallow slope of the OLS relations (see \cref{fig:cHbeta_mfinal}), resulting in a generally stronger extinction even at the lower final masses below \SI{0.55}{\Msun}.

Having been optimized to recover a constant bright-end across a range of masses that is as large as possible, the Optimized PNLFs (purple lines in \cref{fig:pnlf_grid_extinction}) show a PNLF dimming that is right in the middle of the alternative models for younger SSPs.
This leads to a slightly brighter cutoff than $M^*(5007) = -4.5$ for some SSPs, but it quickly approaches a negligible dimming for older SSPs when the lower-mass PNe become intrinsically dimmer. Thus, the Optimized recipe produces the most constant PNLF cutoff across SSPs of different ages and metallicities using the PICS model.

\subsection{Extinctions of de-reddened PNe}
\label{sec:dereddened}

\begin{figure}[!t]
    \begin{center}
    \includegraphics[width=0.9\textwidth]{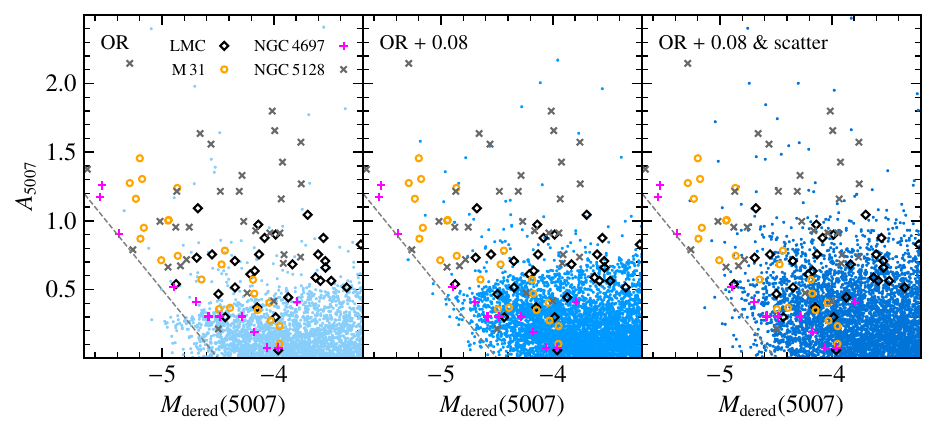}
    \end{center}
    \caption{Circumnebular extinction of \OIII{}, $A_{5007}$, versus the intrinsic de-reddened \OIII{} magnitude, $M_\mathrm{dered}(5007)$, of observed and modeled PNe for three different circumnebular extinction recipes. \textit{Left}: Combined OR. \textit{Middle}: Combined OR with a shift of $\Delta c(\Hbeta) = +0.08$. \textit{Right}: Combined OR with the same shift of +0.08 and an added normally distributed scatter in $c(\Hbeta)$ of $\sigma = 0.1$. The values of the observed PNe are taken from \citet{davis+18}, who observed PNe in the bulge of \m{31} and compiled the data from other works for three other galaxies: the LMC \citep{reid&parker10}, \ngc{4697} \citep{mendez+05}, and \ngc{5128} \citep{walsh+12}. The dashed gray lines indicate the bright-end cutoff of the PNLF of $M(5007) = -4.5$.}
    \label{fig:a5007_m5007}
\end{figure}

Finally, we investigated the relation of intrinsic de-reddened \OIII{} magnitudes with their circumnebular extinctions.
For this, we ran PICS for \num{100000} PNe at an intermediate metallicity of $Z=0.01$ with different circumnebular extinction recipes, \num{1000} PNe each for 100 uniformly spaced ages between 0.25 and \SI{10}{\giga\year}. The relation of their intrinsic \OIII{} magnitudes, $M_\mathrm{dered}(5007)$, with the \OIII{} extinction, $A_{5007}$, is shown in \cref{fig:a5007_m5007}. We also included observations of bright PNe from four nearby galaxies, as compiled by \citet{davis+18}.
These are from \m{31} \citep{davis+18}, the LMC \citep{reid&parker10}, \ngc{4697} \citep{mendez+05}, and \ngc{5128} \citep{walsh+12}.
The dashed line marks the location of the bright-end cutoff for PNe of different extinctions.
In the left and middle panels, the Combined OR and shifted Combined OR recipes were applied, respectively. The elevation to higher extinctions can immediately be seen for the shifted OR recipe: first, the bulk of PNe reach higher values of $A_{5007}$, and second, there are fewer PNe that reside in the regime brighter than the bright-end cutoff (region below the dashed gray line).
Overall, most PNe have lower extinction values of $A_{5007} \lesssim 0.5$ and 0.75 for OR and shifted OR, respectively. This stands in contrast to the observed PNe, where the extinctions of the PNe are more uniformly distributed. Only \ngc{4697} contains a large fraction of PNe with $A_{5007} \lesssim 0.5$, but still has a significant amount of them at higher extinctions (3 out of 12), unlike the modeled PNe.
Additionally, there is a larger fraction of intrinsically bright observed PNe with $M(5007) \lesssim -5.0$ compared to the modeled PNe.

Since the observed extinctions are much greater than the modeled ones in both cases (left and center panel of \cref{fig:a5007_m5007}), we decided to add a random normally distributed scatter to $c(\Hbeta)$ after applying the respective linear relation from \citet{jacoby&ciardullo25}.
The motivation is that any asymmetric nebulae would be expected to also have varying amounts of dust along different lines of sight, thus changing the circumnebular extinction.
We tested an additional Gaussian scatter with $\sigma = 0.1$ applied to the shifted Combined OR relation. This chosen value leads to a slightly larger scatter than is observed in the $c(\Hbeta)$ relation showed in \cref{fig:cHbeta_mfinal}, but which still has the same median behavior as the non-scattered versions.
In the right panel of \cref{fig:a5007_m5007} we show how the $A_\mathrm{5007}$ extinctions are able to reach higher values due to the increased scatter. However, even with this, the large relative amount of high-extinction PNe found in the observations is still not recovered.
Finally, PNe brighter than the bright-end cutoff (those found below the dashed diagonal line) are now on average also further away from the cutoff, with several ending up at magnitudes close to $M(5007) \sim -4.8$.
Therefore, the larger scatter on the one hand increases the extinctions that can generally be reached, but on the other hand also leads to overly bright PNe, though there are also occasional observations of such PNe \citep[e.g.,][fig.~59]{jacoby+24:II}. We conclude that the simple additional scatter for the extinction is not sufficient in bringing the modeled and observed \OIII{} extinctions in agreement.

\section{Discussion}
\label{sec:discussion}

For the \OIII{} circumnebular extinction, $A_{5007}$, the largest effect of the different tested extinction recipes is found at the beginning of the post-AGB phase (\cref{fig:a5007}).
Because only the more massive central stars can come near the bright end within this short time frame due to their faster heating \citep[e.g.,][]{miller_bertolami16}, the only SSP PNLFs that experience significant extinction differences between the recipes are those of the youngest SSPs with $t_\mathrm{age} \lesssim \SI{0.5}{\giga\year}$ (\cref{fig:pnlf_grid_extinction}).
This implies that in particular the early behavior of dust build-up in the nebula is of great importance for the resulting amount of circumnebular extinction.
However, given that the implemented recipes were based on the empirical results of extinction related to final mass from \citet{jacoby&ciardullo25}, we note that the regime of very large extinction is largely beyond the originally fitted range of $c(\Hbeta) \lesssim 0.5$ ($A_{5007} \lesssim 1.2$).
In turn, this range covers most of the PN lifetime after \SI{200}{\year}, where the differences are much smaller as they are more constrained by the observations.

Overall, the modeled PNe have similar extinctions to the observed PNe for the different recipes, but neither the pure OR nor OLS recipes reproduce the same median extinction-final mass relation as the observations (\cref{fig:cHbeta_mfinal}).
This shows that the average extinction relation of the PNe in the brightest magnitude is not the direct desired input relation for the PICS model \citep{valenzuela+25:picsI}, though the slope of the OR recipe is similar to that of the observed median relation.
We addressed this offset by shifting the OR relations up by the respective amount of $\Delta c(\Hbeta) = 0.08$. However, in the future ideally the relation between extinction and final mass will be informed by a larger sample of PNe that directly lie at the bright end to limit the scatter around the fitted relation \citep[see the discussion by][]{jacoby&ciardullo25}.

Despite the Combined OR recipes reaching the second highest extinctions (after the LMC OR recipes; see \cref{fig:a5007}) and generally reproducing the relation between extinction and final stellar mass, it is curious that the models using \emph{Combined OR} do not reach as high extinctions for a full PN population as found in nearby galaxies (\cref{fig:a5007_m5007}).
Even when including a scatter for the extinction that would produce higher values, the distribution still does not match the observations.
There are multiple possible reasons for this discrepancy. For one, the modeled stellar population is artificially created with PNe at uniformly spaced ages and thus does not represent the true star formation history of any of the four observed galaxies.
However, from the model side the only realistic possibility to obtain a similar amount of high-extinction PNe would be to have a much larger fraction of younger stars, which is likely not the case for the observed galaxies.

Furthermore, potential model-intrinsic issues of the recipes used by PICS could lead to PN populations that do not entirely match the observed PNe. While the PNLF properties and many relations like the extinction-final mass relation do agree with observations, the properties of individual PNe could still be systematically offset because of unknown degeneracies.
Another possible explanation is that PICS is missing relevant channels for forming PNe at the bright end, for example accreting white dwarfs \citep[e.g.,][]{soker06, souropanis+23}, blue stragglers \citep[e.g.,][]{ciardullo+05}, and white dwarf mergers \citep[e.g.,][]{yao&quataert23}, which could result in different extinction properties.
These additional possible pathways are supported by the evidence that a significant fraction of PNe are in binary systems \citep{jones&boffin17, jacoby+21, csukai+25}.
Additionally, a random scatter of the final mass through the IFMR as proposed by \citet{jacoby&ciardullo25} could also increase the number of bright PNe with higher extinctions.
Systematic biases in the extinction determination for the observations could lead to a preference of higher extinctions being predicted for the PNe. Then the slope parallel to the bright-end cutoff line seen in \cref{fig:a5007_m5007} for \m{31} and \ngc{4697} could also be the result of selection bias coupled to systematically higher extinctions.

Finally, it is interesting to consider the effects that metallicity would have on the extinction, which are not accounted for in the tested recipes. At higher metallicity, there should be more metals to contribute to the formation of dust for equal stellar masses.
As also pointed out by \citet{jacoby&ciardullo25}, curiously their derived relations between extinction and final mass show no significant differences between \m{31} and the LMC PNe, even though the PNe in the bulge of \m{31} have roughly twice the metal mass fraction as those in the LMC.
The LMC PNe even seem to reach higher extinctions at a given final mass (\cref{fig:cHbeta_mfinal}), though this seems to stand in contrast to the overall lower measured \OIII{} extinctions \citep{davis+18}, seen in \cref{fig:a5007_m5007}.

To conclude, while circumnebular extinction surely plays a central role in limiting the \OIII{} magnitude of PNe in intermediate-to-younger stellar populations to the universal bright-end cutoff, the details around its dependency on other physical properties are not straightforward.
While recipe variations in the PICS model affect the youngest stellar populations the most due to being the least constrained observationally, the general discrepancies between the modeled extinctions and those determined from observations make it very difficult to disambiguate between the recipes.
Further observations and advances on the understanding of circumnebular dust evolution will be necessary to improve the modeling.

\section*{Conflict of Interest Statement}

The authors declare that the research was conducted in the absence of any commercial or financial relationships that could be construed as a potential conflict of interest.

\section*{Author Contributions}

LMV developed the project idea, performed the analysis and wrote the original draft. GHJ and RSR participated in the conceptualization of the project and provided comments on the manuscript. MMMB and RHM provided comments on the manuscript.

\section*{Funding}
LMV acknowledges support by the German Academic Scholarship Foundation (Studienstiftung des deutschen Volkes) and the Marianne-Plehn-Program of the Elite Network of Bavaria.

\section*{Acknowledgments}
We thank the two referees for their comments that helped improve the manuscript.

\section*{Data Availability Statement}
The original contributions presented in the study are included in the article/supplementary material, further inquiries can be directed to the corresponding author/s.

\bibliographystyle{Frontiers-Harvard} 
\bibliography{bib}

\begin{thebibliography}{}
\makeatletter
\relax
\def\mn@urlcharsother{\let\do\@makeother \do\$\do\&\do\#\do\^\do\_\do\%\do\~}
\def\mn@doi{\begingroup\mn@urlcharsother \@ifnextchar [ {\mn@doi@} {\mn@doi@[]}}
\def\mn@doi@[#1]#2{\def\@tempa{#1}\ifx\@tempa\@empty \href {http://dx.doi.org/#2} {doi:#2}\else \href {http://dx.doi.org/#2} {#1}\fi \endgroup}
\def\mn@eprint#1#2{\mn@eprint@#1:#2::\@nil}
\def\mn@eprint@arXiv#1{\href {http://arxiv.org/abs/#1} {{\tt arXiv:#1}}}
\def\mn@eprint@dblp#1{\href {http://dblp.uni-trier.de/rec/bibtex/#1.xml} {dblp:#1}}
\def\mn@eprint@#1:#2:#3:#4\@nil{\def\@tempa {#1}\def\@tempb {#2}\def\@tempc {#3}\ifx \@tempc \@empty \let \@tempc \@tempb \let \@tempb \@tempa \fi \ifx \@tempb \@empty \def\@tempb {arXiv}\fi \@ifundefined {mn@eprint@\@tempb}{\@tempb:\@tempc}{\expandafter \expandafter \csname mn@eprint@\@tempb\endcsname \expandafter{\@tempc}}}

\bibitem[\protect\citeauthoryear{Balick et~al.,}{Balick et~al.}{2013}]{balick+13}
Balick B.,  et~al., 2013, \mn@doi [\apj] {10.1088/0004-637X/774/1/3}, \href {https://ui.adsabs.harvard.edu/abs/2013ApJ...774....3B} {774, 3}

\bibitem[\protect\citeauthoryear{Ciardullo}{Ciardullo}{2012}]{ciardullo12}
Ciardullo R.,  2012, \mn@doi [\apss] {10.1007/s10509-012-1061-2}, \href {https://ui.adsabs.harvard.edu/abs/2012Ap&SS.341..151C/abstract} {341, 151}

\bibitem[\protect\citeauthoryear{Ciardullo}{Ciardullo}{2022}]{ciardullo22}
Ciardullo R.,  2022, \mn@doi [Frontiers in Astronomy and Space Sciences] {10.3389/fspas.2022.896326}, \href {https://ui.adsabs.harvard.edu/abs/2022FrASS...9.6326C} {9, 896326}

\bibitem[\protect\citeauthoryear{Ciardullo \& Jacoby}{Ciardullo \& Jacoby}{1999}]{ciardullo&jacoby99}
Ciardullo R.,  Jacoby G.~H.,  1999, \mn@doi [\apj] {10.1086/307025}, \href {https://ui.adsabs.harvard.edu/abs/1999ApJ...515..191C} {515, 191}

\bibitem[\protect\citeauthoryear{Ciardullo et~al.,}{Ciardullo et~al.}{2002}]{ciardullo+02:pnlfXII}
Ciardullo R.,  et~al., 2002, \mn@doi [\apj] {10.1086/342180}, \href {https://ui.adsabs.harvard.edu/abs/2002ApJ...577...31C} {577, 31}

\bibitem[\protect\citeauthoryear{Ciardullo et~al.,}{Ciardullo et~al.}{2005}]{ciardullo+05}
Ciardullo R.,  et~al., 2005, \mn@doi [\apj] {10.1086/431353}, \href {https://ui.adsabs.harvard.edu/abs/2005ApJ...629..499C} {629, 499}

\bibitem[\protect\citeauthoryear{{Csukai} et~al.,}{{Csukai} et~al.}{2025}]{csukai+25}
{Csukai} A.,  et~al., 2025, \mn@doi [\mnras] {10.1093/mnras/staf1552}, \href {https://ui.adsabs.harvard.edu/abs/2025MNRAS.543.3035C} {543, 3035}

\bibitem[\protect\citeauthoryear{Davis et~al.,}{Davis et~al.}{2018}]{davis+18}
Davis B.~D.,  et~al., 2018, \mn@doi [\apj] {10.3847/1538-4357/aad3c4}, \href {https://ui.adsabs.harvard.edu/abs/2018ApJ...863..189D} {863, 189}

\bibitem[\protect\citeauthoryear{Dopita \& Meatheringham}{Dopita \& Meatheringham}{1991}]{dopita&meatheringham91:II}
Dopita M.~A.,  Meatheringham S.~J.,  1991, \mn@doi [\apj] {10.1086/170377}, \href {https://ui.adsabs.harvard.edu/abs/1991ApJ...377..480D} {377, 480}

\bibitem[\protect\citeauthoryear{Dopita et~al.,}{Dopita et~al.}{1992}]{dopita+92}
Dopita M.~A.,  et~al., 1992, \mn@doi [\apj] {10.1086/171186}, \href {https://ui.adsabs.harvard.edu/abs/1992ApJ...389...27D} {389, 27}

\bibitem[\protect\citeauthoryear{Jacoby}{Jacoby}{1989}]{jacoby89:pnlfI}
Jacoby G.~H.,  1989, \mn@doi [\apj] {10.1086/167274}, \href {https://ui.adsabs.harvard.edu/abs/1989ApJ...339...39J} {339, 39}

\bibitem[\protect\citeauthoryear{Jacoby \& Ciardullo}{Jacoby \& Ciardullo}{2025}]{jacoby&ciardullo25}
Jacoby G.~H.,  Ciardullo R.,  2025, \mn@doi [\apj] {10.3847/1538-4357/adc0fb}, \href {https://ui.adsabs.harvard.edu/abs/2025ApJ...983..129J} {983, 129}

\bibitem[\protect\citeauthoryear{Jacoby et~al.,}{Jacoby et~al.}{2021}]{jacoby+21}
Jacoby G.~H.,  et~al., 2021, \mn@doi [\mnras] {10.1093/mnras/stab2045}, \href {https://ui.adsabs.harvard.edu/abs/2021MNRAS.506.5223J} {506, 5223}

\bibitem[\protect\citeauthoryear{Jacoby et~al.,}{Jacoby et~al.}{2024}]{jacoby+24:II}
Jacoby G.~H.,  et~al., 2024, \mn@doi [\apjs] {10.3847/1538-4365/ad2166}, \href {https://ui.adsabs.harvard.edu/abs/2024ApJS..271...40J} {271, 40}

\bibitem[\protect\citeauthoryear{Jones \& Boffin}{Jones \& Boffin}{2017}]{jones&boffin17}
Jones D.,  Boffin H. M.~J.,  2017, \mn@doi [\nat] {10.1038/s41550-017-0117}, \href {https://ui.adsabs.harvard.edu/abs/2017NatAs...1E.117J} {1, 0117}

\bibitem[\protect\citeauthoryear{M{\'e}ndez et~al.,}{M{\'e}ndez et~al.}{2005}]{mendez+05}
M{\'e}ndez R.~H.,  et~al., 2005, \mn@doi [\apj] {10.1086/430498}, \href {https://ui.adsabs.harvard.edu/abs/2005ApJ...627..767M} {627, 767}

\bibitem[\protect\citeauthoryear{Miller~Bertolami}{Miller~Bertolami}{2016}]{miller_bertolami16}
Miller~Bertolami M.~M.,  2016, \mn@doi [\aap] {10.1051/0004-6361/201526577}, \href {https://ui.adsabs.harvard.edu/abs/2016A&A...588A..25M/abstract} {588, A25}

\bibitem[\protect\citeauthoryear{Reid \& Parker}{Reid \& Parker}{2010}]{reid&parker10}
Reid W.~A.,  Parker Q.~A.,  2010, \mn@doi [\mnras] {10.1111/j.1365-2966.2010.16635.x}, \href {https://ui.adsabs.harvard.edu/abs/2010MNRAS.405.1349R} {405, 1349}

\bibitem[\protect\citeauthoryear{Rekola et~al.,}{Rekola et~al.}{2005}]{rekola+05}
Rekola R.,  et~al., 2005, \mn@doi [\mnras] {10.1111/j.1365-2966.2005.09166.x}, \href {https://ui.adsabs.harvard.edu/abs/2005MNRAS.361..330R} {361, 330}

\bibitem[\protect\citeauthoryear{Sarzi et~al.,}{Sarzi et~al.}{2011}]{sarzi+11}
Sarzi M.,  et~al., 2011, \mn@doi [\mnras] {10.1111/j.1365-2966.2011.18900.x}, \href {https://ui.adsabs.harvard.edu/abs/2011MNRAS.415.2832S} {415, 2832}

\bibitem[\protect\citeauthoryear{Soker}{Soker}{2006}]{soker06}
Soker N.,  2006, \mn@doi [\apj] {10.1086/500291}, \href {https://ui.adsabs.harvard.edu/abs/2006ApJ...640..966S} {640, 966}

\bibitem[\protect\citeauthoryear{Souropanis et~al.,}{Souropanis et~al.}{2023}]{souropanis+23}
Souropanis D.,  et~al., 2023, \mn@doi [\mnras] {10.1093/mnras/stad564}, \href {https://ui.adsabs.harvard.edu/abs/2023MNRAS.521.1808S} {521, 1808}

\bibitem[\protect\citeauthoryear{Valenzuela et~al.,}{Valenzuela et~al.}{2019}]{valenzuela+19}
Valenzuela L.~M.,  et~al., 2019, \mn@doi [\apj] {10.3847/1538-4357/ab4e96}, \href {https://ui.adsabs.harvard.edu/abs/2019ApJ...887...65V} {887, 65}

\bibitem[\protect\citeauthoryear{{Valenzuela} et~al.,}{{Valenzuela} et~al.}{2025a}]{valenzuela+25:pics}
{Valenzuela} L.~M.,  et~al., 2025a, in {De Marco} O.,  et~al., eds,  IAU Symposium Vol. 384, Planetary Nebulae: A Universal Toolbox in the Era of Precision Astrophysics. pp 69--75 (\mn@eprint {arXiv} {2412.08702}), \mn@doi{10.1017/S1743921323005458}

\bibitem[\protect\citeauthoryear{{Valenzuela} et~al.,}{{Valenzuela} et~al.}{2025b}]{valenzuela+25:picsI}
{Valenzuela} L.~M.,  et~al., 2025b, \mn@doi [\aap] {10.1051/0004-6361/202553974}, \href {https://ui.adsabs.harvard.edu/abs/2025A&A...699A.371V} {699, A371}

\bibitem[\protect\citeauthoryear{Ventura et~al.,}{Ventura et~al.}{2014}]{ventura+14}
Ventura P.,  et~al., 2014, \mn@doi [\mnras] {10.1093/mnras/stu028}, \href {https://ui.adsabs.harvard.edu/abs/2014MNRAS.439..977V} {439, 977}

\bibitem[\protect\citeauthoryear{Walsh et~al.,}{Walsh et~al.}{2012}]{walsh+12}
Walsh J.~R.,  et~al., 2012, \mn@doi [\aap] {10.1051/0004-6361/201118580}, \href {https://ui.adsabs.harvard.edu/abs/2012A&A...544A..70W} {544, A70}

\bibitem[\protect\citeauthoryear{Yao \& Quataert}{Yao \& Quataert}{2023}]{yao&quataert23}
Yao P.~Z.,  Quataert E.,  2023, \mn@doi [\apj] {10.3847/1538-4357/acfed9}, \href {https://ui.adsabs.harvard.edu/abs/2023ApJ...957...30Y} {957, 30}

\makeatother
\end{thebibliography}





\end{document}